\documentclass{article}
\usepackage{graphicx}
\usepackage{subfiles}
\usepackage{multirow}
\usepackage{url}
\usepackage{enumitem}
\usepackage{fancyhdr}
\usepackage{a4wide}

\begin{document}
\title{On the Role of Similarity in Detecting Masquerading Files}
%
%
\author{
Jonathan Oliver\textsuperscript{1 2}		(jonathano@vmware.com)	\\
Jue Mo\textsuperscript{1}			(mjue@vmware.com)	\\
Susmit Yenkar\textsuperscript{1}		ysusmit@vmware.com	\\
Raghav Batta\textsuperscript{1}			rbatta@vmware.com	\\
Sekhar Josyoula\textsuperscript{1}		(sjosyoula@vmware.com)	\\
	\\
	1. VMware, Palo Alto, CA 94304, USA. \\
	2. University of Queensland, Queensland, Australia.
}
\maketitle              

\begin{abstract}
Similarity has been applied to a wide range of security applications, typically used in machine learning models.
We examine the problem posed by masquerading samples; that is samples crafted by bad actors to be similar or near identical to legitimate samples.
We find that these samples potentially create significant problems for machine learning solutions.
The primary problem being that bad actors can circumvent machine learning solutions by using masquerading samples.

We then examine the interplay between digital signatures and machine learning solutions.
In particular, we focus on executable files and code signing.
We offer a taxonomy for masquerading files.
We use a combination of similarity and clustering to find masquerading files.
We use the insights gathered in this process to offer improvements to similarity based and machine learning security solutions.
\end{abstract}



\section{Introduction}
\label{sec:intro}

Similarity has a wide range of uses in computer security.
We can determine that a security object is similar to a known good entity or to a known bad entity.
Similarity has been applied to a wide range of security objects including applications in
spam filters, detecting phishing pages and detecting malware variants.

We use similarity hashes as our approach for similarity as SSDEEP \cite{kornblum2006identifying} and TLSH \cite{oliver2013TLSH}
are available in public repositories of malware \cite{MalwareBazaar} and
have been adopted by the STIX and MISP standards \cite{stix,misp}.
Similarity hashes are typically combined with clustering approaches to form a cluster model of the data \cite{wallace2015optimizing,oliver2020hac}.
An idealized framework for using similarity and clustering in security works as follows:
\begin{enumerate}[label=(\alph*)]
	\item A training set of (partially) labeled data is clustered according to the distance function.
	\item A reputation is assigned to each cluster based on its membership.
	\item New security objects can be classified by finding the most similar cluster according to the distance function being employed\footnote{
		Issues of scale can be address by tools such as Approximate Nearest Neighbor (ANN) \cite{ann-benchmarks,annoy}.}.
		The naive approach for using such a machine learning solution is to make these predictions by selecting the
		most common category from the closest cluster.
\end{enumerate}

For most problems in the computer security domain the above process suffers from a fundamental problem.
Let us assume that the clustering worked as intended and created a clear separation of the clusters.
Bad actors are known to use a wide range of tricks to make malicious content similar to legitimate content \cite{MitreMasquerading}.
We shall term such a malicious sample as being a {\em masquerading sample}.
Some small number of clusters will contain members created by bad actors which are masquerading as legitimate samples.

The masquerading samples potentially cause serious problems for machine learning solutions.
This is made worse by the fact that security datasets are typically only partially labeled \cite{CIOArticle}, so
we expect many clusters to have members which are missing labels and some masquerading samples to be un-labeled.
The problems arise when a masquerading sample is included in an otherwise legitimate cluster.
Some of the problems include:
\begin{enumerate}
	\item Bad actors can circumvent machine learning solutions by using masquerading samples.
		The naive approach for prediction will assign a legitimate category for a masquerading sample.
	\item A cluster with masquerading members may be labeled as malicious, resulting in false positive predictions.
	\item If a cluster with undetected masquerading members is used to form the baseline of legitimate activity,
		then the baseline will include malicious behaviours.
\end{enumerate}
Problem 1 is the most serious issue. We note that it only applies if we use the naive approach for prediction.
An extreme solution would be to configure the machine learning to avoid making predictions for clusters which contain both
masquerading members and legitimate members.
This extreme solution is unsatisfactory as it means that the model will fail to make predictions for important clusters\footnote{
We will see examples of files masquerading as Microsoft and Google in the Section~\ref{sec:collectingMasqueradingFiles}.}.

Traditional security employs digital signatures for applications such as signing executable programs \cite{wikiCodeSigning}
and authenticating email \cite{wikiEmailAuthentication} to detect malicious attempts at masquerading.
In this paper, we explore the issue of the interplay of machine learning with similarity-based methods and digital signatures.
While there is a wealth of academic literature on both sides of this interplay,
there is virtually no discussion on how these technologies should be applied together to produce results.

To explore this issue, we focus on executable files and concentrate on the interplay of code signing with machine learning.
The contributions of this paper are:
\begin{itemize}
	\item We give a taxonomy for masquerading files.
	\item We show how a similarity-based system can be used to find real world masquerading files in a public database of malware.
		While these real-world masquerading samples are in the minority, they confirm the problems raised here.
	\item We offer methods for including digital signatures in ML and similarity-based solutions.
\end{itemize}


\section{A Taxonomy of Masquerading Files}
\label{sec:taxonomy}

There is a range of methods available to bad actors for creating masquerading files.
The Mitre Attack Framework \cite{MitreAttackFramework}
includes 7 types of masquerading under T1036 \cite{MitreMasquerading} of which 6 are relevant to executable files.

\begin{itemize}
	\item Syntactic Masquerading: These attacks focus on tricking end-users that the file is legitimate. They include
		renaming files to legitimate filenames (Mitre T1036.003),
		obfuscating filenames with spaces (T1036.006), and
		obfuscating filenames with double file extensions (T1036.007).
		These methods may also evade security rules and methods which rely on surface level features (such as filenames)
		or when digital signatures are not available (for example with unsigned files).
	\item Content Masquerading: This involves inserting malicious content into an existing file (often signed).
		When malicious content is inserted into a signed file, Mitre terms this "Invalid Code Signature" (T1036.001).
		The Neshta malware family takes existing executable files and inserts malicious content \cite{Neshta}.
		If the file was signed, then it will leave the original X509 certificate in the executable and the new file will
		no longer have a valid signature.
	\item Certificate Attacks: This is the situation where an attacker compromises the digital signature \cite{leavitt2011internet}
		or the part of the OS that checks the digital signature \cite{HijackingDigitalSignatures}.
	\item Supply Chain Attacks: This is the situation where an attacker compromises the source of software \cite{lella2021enisa,martinez2021software,barr2022exorcist}.
	\item Adversarial ML: This is an area of Machine Learning where a sample is created to trick a
		ML system into making an incorrect classification.
		This approach has proven effective at tricking image recognition systems \cite{szegedy2013intriguing,kurakin2018adversarial}
		and has proven effective for deceiving a ML security solution on executable files \cite{ashkenazy2019cylance}.
\end{itemize}
For the remainder of the paper, we focus on content masquerading, certificate attacks and supply chain attacks.
We will not study Syntactic Masquerading files as they are often self-evident when inspected and would typically be detected by
antispam and antivirus products.
At this stage, we have seen no evidence that adversarial ML has been adopted by cybercriminals for the creation of malware.


\section{Collecting Masquerading Files}
\label{sec:collectingMasqueradingFiles}

The first step in performing an analysis of masquerading files is to collect a dataset of them.

One important class of masquerading files is files where the binary content of the file is similar to legitimate software
and the file has been classified as malicious.
We can use the Malware Bazaar \cite{MalwareBazaar} dataset which as of August 2023 has over 700,000 malware samples.
It is ideal for research purposes such as this as
(i) it is a malware dataset of reasonable size,
(ii) all submissions are TLP:WHITE,
(iii) it has been indexed by hashes that enable similarity search (TLSH and SSDEEP), and
(iv) all files are available for downloading by other researchers.

We applied the following process to get a candidate list of masquerading files:
\begin{enumerate}
	\item Build a cluster model of a large set of customer files.
		This cluster model is assumed to contain a significant majority of legitimate files.
		We built this model using the HAC-T algorithm with the TLSH distance function \cite{oliver2020hac}.
	\item Process the Malware Bazaar dataset of malware using the TLSH distance function and identify those files that
		were within a distance of 30 of a customer-based cluster (most likely legitimate).
		The distance 30 was used as the threshold to target a false positive rate of $0.002 \%$ (see Table 2 of \cite{oliver2013TLSH})
		which would reflect an estimate for the probability of files being incorrectly assigned to a cluster.
\end{enumerate}
The process resulted in 703 of the 700,000 malware samples being flagged as candidate masquerading files.

Table~\ref{TableSignState} shows the number of files with each signature state.
We then categorized the state of the certificate according to a combination of
the files signing state, metadata provided by Malware Bazaar and some crowdsourced rules
(including "cert\_blocklist" from \\ https://github.com/reversinglabs/reversinglabs-yara-rules and
"knownbad\_certs" from \\ https://github.com/ditekshen/detection).


\begin{table}[!t]
    \caption{The number of files with each signature state.}
    \label{TableSignState}
    \centering
    \begin{tabular}{|l|l|}
    \hline
	Signature State	& Count \\
    \hline
	signed: certificate chain could not be built to a trusted root authority	& 2	\\
	signed: using stolen or revoked certificates					& 4	\\
	signed: certificate not in validity period					& 7	\\
	signed: certificate used for digitally signing malware				& 10	\\
	signed: certificate revoked							& 11	\\
	signed: verified								& 38	\\
	not signed: contains x509 certificate						& 44	\\
	signed: not verified								& 94	\\
	not signed: no signature							& 493	\\
    \hline
    \end{tabular}
\end{table}

We now go through the various cases and give clear cut example where we point back to information
on Malware Bazaar. We highlight how analysis with similarity is helping understand the nature of masquerading files.
The examples below were created by hand to determine that the files are masquerading.
We note that each case here could be automated in a straightforward way to generate alerts or detections for the case in question.
Any alerts or detections generated can give reasoning as to the nature of the anomaly discovered.

We believe that the determination that a file is a masquerading file would be very difficult without the additional cluster information.
We note that security analyses of files very rarely specify which legitimate file a malicious file is masquerading as.

\subsection{The No Signature Case}
 
For each of these cases, we will show a table with the sample from Malware Bazaar listed under "File" and
details of the closest cluster which includes details of a member.
The first example we examined was an example where the malicious sample had no signature, while the cluster
had legitimate signed files. \\
\vspace{4pt}
\begin{tabular}{|l|l|}
\hline \multicolumn{2}{|c|}{File} \\ \hline
	SHA256 &
		27aa6523fdb14ef7bc83fcfd2d28c752fb8984acbe4d3fa0550ee36ace16bf77
	\\ TLSH & {\scriptsize
		T197248C2032C0C073C062147641B5C7F55EBB78755A66AA8BABCB1FB94F252D2E72938D
	} \\ Filename &
		RemovePillow.exe
	\\ Notes &
		10 vendors detected, collected from web-download
\\ \hline \multicolumn{2}{|c|}{Closest Cluster} \\ \hline
	TLSH & {\scriptsize
		T100148C2072C0C073C063147641B5C7F55EBB78755A65AA8BBBCB5FB90F252D2E72938D
	} \\ SHA256	&
		b7b36285a5249d63f65697ea598bcb98ecf10677a5ba04a8425e540c9cbcc5e1
	\\ Filename &
		wininst-9.0.exe
	\\ Signer &
		Corel Corporation
\\ \hline \multicolumn{2}{|c|}{Distance from File to Closest Cluster: 8}
\\ \hline
\end{tabular}
\vspace{4pt}

We note that similarity has been helpful in identifying what legitimate library the malware is mimicking.
Both the malicious and legitimate files in this association are related to the installing / uninstalling of Python environments and
in particular the Pillow Python library.


\subsection{The Not Verified Case}

We next examine is the "not verified" case.
This is where a signature has been left in a file, but the verification process fails in some way.
This can occur when a bad actor (or another piece of malware) has modified an existing signed file to add malicious code.
Typically, the digital signature will not verify because the file no longer satisfies the checksum in the signature.

Our process found the following example: \\
\vspace{4pt}
\begin{tabular}{|l|l|}
\hline \multicolumn{2}{|c|}{File} \\ \hline
	SHA256 &
		e002c57c0bf40d4f51f798ee07d6440cd1b68f30696cc29980e51cfced68c595
	\\ TLSH & {\scriptsize
		T124067C87E1E221DCC17B803486AB9713F671385923109AF797C0EA353A37FD06576BA6
	} \\ Filename &
		AIDE.dll
	\\ Signer &
		Adobe (fails to verify)
\\ \hline \multicolumn{2}{|c|}{Closest Cluster} \\ \hline
	TLSH & {\scriptsize
		T100067D87E1E221DCC17B803486AB9713FA71385923109AF797C0EA353A37FD06576B96
	} \\ SHA256	&
		70ee341243edb68d3e1ee6100a96a859212340a72d1dceab93e65de56856ed7b
	\\ Filename &
		AIDE.dll
	\\ Signer &
		Adobe
\\ \hline \multicolumn{2}{|c|}{Distance from File to Closest Cluster: 4}
\\ \hline
\end{tabular}
\vspace{4pt}

Again, similarity was useful in determining that a file had a serious anomaly that could be explained from a security perspective.
Even if the digital signature had been stripped all together from the malicious file, then we can still raise an anomaly for
files near this cluster, namely that we expect them to be signed by Adobe.

\subsection{The Contains a X509 Certificate Case}

We now cover the "contains x509 certificate" case.
In this situation, the file has had more of the digital certificate removed, but the X509 certificate remains in the file.
Again, the digital signature should not verify.

Our process found the following example: \\
\vspace{4pt}
\begin{tabular}{|l|l|}
\hline \multicolumn{2}{|c|}{File} \\ \hline
	SHA256 &
		f1a4bbcf6335464a249378fc07e95c8f2f5315b9f8b6e2b845f2317894e80f56
	\\ TLSH & {\scriptsize
		T18C14AE21B180D072E627147186A8CEB109BA7C7A5AB0444F7BED3A791F737E0426D79F
	} \\ Filename &
		java.exe
	\\ Notes &
		The file contains the remains of a X509 certificate.
	\\  &
		Listed as having 16 vendor detections.
\\ \hline \multicolumn{2}{|c|}{Closest Cluster} \\ \hline
	TLSH & {\scriptsize
		T100049D61B180D072E567047189A8CEB04AB67C7A59B0844F7BED76790FB33E1826979F
	} \\ SHA256	&
		2e83f8904ea9744207d4128c6e0f3578dbbe41e197b159f9659a97740209f102
	\\ Filename &
		java.exe
	\\ Signer &
		Oracle
\\ \hline \multicolumn{2}{|c|}{Distance from File to Closest Cluster: 24}
\\ \hline
\end{tabular}
\vspace{4pt}

Similarity was useful in determining that a file had an anomaly and explaining that anomaly.
Even if the X509 certificate had been completely stripped from the file, then we can still determine that
the file has an anomaly requiring explanation.
The anomaly is that java.exe files near this cluster should be signed by "Oracle".

\subsection{The Certificate Revoked Case}

We now cover the "certificate revoked" case.
In this situation, the file has been digitally signed by a certificate which has been revoked by the certificate issuer.

Our process found the following example: \\
\vspace{4pt}
\begin{tabular}{|l|l|}
\hline \multicolumn{2}{|c|}{File} \\ \hline
	SHA256 &
		ea2decec34ae3129d5da1f2035b34cff3c9f656bb4423904ef6b0a3ca5f47d5e
	\\ TLSH & {\scriptsize
		T1055549716142D273D063417DDD64E6F7546BFDB9CB60A4E722887E2E3A303C22A3196B
	} \\ Filename &
		TeamViewer\_Note.exe
	\\ Signer &
		Hartex LLC (this certificate has been revoked)
	\\ Notes &
		Malware Bazaar lists this file as having a code signing certificate
	\\  &
		which has been used for signing other malware.
\\ \hline \multicolumn{2}{|c|}{Closest Cluster} \\ \hline
	TLSH & {\scriptsize
		T1005539B17282D233D463007CD964D6F6506BFDB4CB60A4EB62D87E2E39303C12A3596B
	} \\ SHA256	&
		14ab8a0258245ebe88222c9bcd8c29ddfade2cc52dcdd7ffcb1a171d0c7a51e4
	\\ Filename &
		TeamViewer\_Note.exe
	\\ Signer &
		TeamViewer GmbH
\\ \hline \multicolumn{2}{|c|}{Distance from File to Closest Cluster: 24}
\\ \hline
\end{tabular}
\vspace{4pt}

Once again, similarity was useful in determining that a file had an anomaly (inconsistency in the Signer),
which warranted further investigation of the certificate.

\subsection{The Certificate Used for Signing Malware Case}

We now cover the "certificate used for digitally signing malware" case.
In this situation, the file has been digitally signed by a certificate which has a history of signing other malware samples.

Our process found a set of 
files that matched the cert\_blocklist ruleset at \\ https://github.com/reversinglabs/reversinglabs-yara-rules.
8 files on Malware Bazaar are all similar to the cluster centered at
\begin{verbatim}
    T10005522A56D8B969E3F69B307FF252D3BB69BC523834CC0E11D5030D0969A42FDA076E
\end{verbatim}
The SHA256 of the 8 files are:
\begin{verbatim}
    274a1df26f7cc09917dfcc151e26d20778a81408f959e7ff36823727e248f015
    7f2094eb1534c20b79bb44b68ff5b0126d3ce3401b409569037aeac022b139de
    fbe5392d8a99a75efd085f6f23d19290d1c2febbf22def3e98687cf53672d6ac
    f2f8e9fcac33f0a957f825522bc4fc43348e00adf6b534f14ccf30f44a5b86c7
    dd23e25e33025599a7df947b96e8b00c2348f3c0a8901b9593ef98b0fd30c94c
    d658764009896365bfc6c896a1f242b344c58e06ca4007b5d98fc48df26bfa69
    c896094017e57358eafffebf3f373d60a238739184ff8a9d43bd08469f752d7a
    54eace7780d77504a6a87991a01ea130f6cdf33b45b54ff6fd83736432092afc
\end{verbatim}
Interestingly this is a situation where the apparent legitimate members of the cluster
(such as setup.exe with a copyright notice by Microsoft \\
167cb9d4bedd8c92cecc8ca8cc658034f7759cd5ef1560fea558278e3a0ced27)
are not signed.




\subsection{The No Trusted Root Authority Case}
We now consider the case "signed: certificate chain could not be built to a trusted root authority".
This case occurs when the root authority / certificate chain has a record with a lack of trust or multiple reported malware.

Our process found the following example: \\
\vspace{4pt}
\begin{tabular}{|l|l|}
\hline \multicolumn{2}{|c|}{File} \\ \hline
	SHA256 &
		bec327afe49789c484820e4b1c1e477d8e7a3d0134b5a9691d05d9d7cb317f11
	\\ TLSH & {\scriptsize
		T1B3831F9D366072EFC857D4729EA86CA4EB5074BB831F4213A02715ADEE4D89BDF140F2
	} \\ Filename &
		Update\_Service\_ALTDNS.exe
	\\ Signer &
		Global Alt Network Soft Certification
	\\ Notes &
		Malware Bazaar lists this file as having a certificate chain
	\\  &
		which has used for multiple reported malware.
\\ \hline \multicolumn{2}{|c|}{Closest Cluster} \\ \hline
	TLSH & {\scriptsize
		T100932E9D762072EFC857C472DEA82C68EA6075BB831F4203902715EDAE4D997CF140F2
	} \\ SHA256	&
		023cfe93385d9b8aa13a1f5a257da627f908f1c7fa6cf88dfeae95b92cc5061f
	\\ Filename &
		LscShim.exe
	\\ Signer &
		LENOVO (UNITED STATES) INC.
\\ \hline \multicolumn{2}{|c|}{Distance from File to Closest Cluster: 21}
\\ \hline
\end{tabular}
\vspace{4pt}

\section{Finding Masquerading Files in Clusters}
\label{sec:FindingMasqueradingFilesInlusters}

Another approach for finding masquerading files is to find clusters with inconsistent signature information.
We searched through the model for clusters where:
\begin{itemize}
	\item We had a majority that were signed and verified.
	\item We had a minority that were unsigned.
\end{itemize}
This search process found the following example: \\
\vspace{4pt}
\begin{tabular}{|l|l|}
\hline \multicolumn{2}{|c|}{Unsigned File} \\ \hline
	SHA256 &
		0a1cfbf61797a565b649d443dcf6102f0e179b68e2582a788a97acf05a4ecd72
	\\ TLSH & {\scriptsize
		T18E75AE05F951D07AC1162070E41DF3396B345E59CB214ADFE7D87E9A3EB02D12A3A2AF
	} \\ Filename &
		chrome.exe
	\\ Signer &
		Google LLC (but does not pass signature verification)
	\\ Notes &
		This file has a X509 certificate claiming it is signed by "Google LLC"
	\\  &
		but does not pass signature verification. This file is a version
	\\  &
		of Chrome which was corrupted by the Neshta malware family \cite{Neshta}.
\\ \hline \multicolumn{2}{|c|}{Cluster and Signed Member Details} \\ \hline
	TLSH & {\scriptsize
		T10075AE01F850D0B6D5122071F41DF339AA355E198B658EDBE3987E9A3FB02D25A3A39F
	} \\ SHA256	&
		606394cc56a3f5b37dff0540795823be3a78f58dace822855f5557ce1ebb4ffe
	\\ Filename &
		chrome.exe
	\\ Signer &
		Google LLC
	\\ Notes &
		This cluster has 45 members all of which were chrome.exe
	\\ &
		signed by "Google LLC".
\\ \hline \multicolumn{2}{|c|}{Distance from File to Closest Cluster: 29}
\\ \hline
\end{tabular}
\vspace{4pt}

This property of clusters mostly containing legitimate signed files with a small number of unsigned versions is an indicator
that the unsigned files may be infected with Neshta.

\section{Clusters Related to Supply Chain Attacks}
\label{sec:ClustersRelatedToSupplyChainAttacks}

Another cluster property to investigate is clusters where:
\begin{itemize}
	\item The members are signed and verified.
	\item The members differ in file reputation.
\end{itemize}
This search process found the following example: \\
\vspace{4pt}
\begin{tabular}{|l|l|}
\hline \multicolumn{2}{|c|}{Bad Reputation Files} \\ \hline
	SHA256 &
		ce77d116a074dab7a22a0fd4f2c1ab475f16eec42e1ded3c0b0aa8211fe858d6
	\\ TLSH & {\scriptsize
		T1EA25C60177EC8A09E1FF2B75AAB441280B73F95A9A76D75E294C109E0FB3B008E51777
	} \\ SHA256 &
		019085a76ba7126fff22770d71bd901c325fc68ac55aa743327984e89f4b0134
	\\ TLSH & {\scriptsize
		T14F25C60177EC8A09E1FF2B75AAB441280B73F95A9A76D75E194C109E0FB3B008E517B7
	} \\ SHA256 &
		32519b85c0b422e4656de6e6c41878e95fd95026267daab4215ee59c107d6c77
	\\ TLSH & {\scriptsize
		T1F525C50173EC8A49F5FF2B74AAB441680B73B8569A7AD74D154C619E0FB3B008E11BB7
	} \\ Filename &
		SolarWinds.Orion.Core.BusinessLayer.dll
	\\ Signer &
		Solarwinds Worldwide, LLC
\\ \hline \multicolumn{2}{|c|}{Cluster and Good Reputation Member} \\ \hline
	TLSH & {\scriptsize
		T10025D54177FC4A09F6FE2B74AAB441190B73B91AAA7AD74E154C209E0FB3B40CE617B7
	} \\ SHA256	&
		671d9bd3ef6b0ecb0507dda84ed5800238844b4deec5f387ff503835909a8cee
	\\ Filename &
		SolarWinds.Orion.Core.BusinessLayer.dll
	\\ Signer &
		Solarwinds Worldwide, LLC
\\ \hline \multicolumn{2}{|c|}{Distances from Files to Closest Cluster: 25, 23, 24}
\\ \hline
\end{tabular}
\vspace{4pt}

The malicious files are a part of the Solarwinds supply chain attack \cite{martinez2021software}.
At the time of the attack using similarity on file content would not be useful to detect this attack.
We do note that having the cluster of legitimate files could help establish a baseline for the behaviour of the files within the cluster
which may be useful for identifying anomalous behavior during incidence response.


\section{Conclusion and Future Work}\label{sec:conclusion_future_work}

We have seen that a sample of candidate list of 703 files from Malware Bazaar indeed includes
masquerading files
with a range of sophistication levels.
This would indicate that we expect at least 1 in 1000 malware is a masquerading file\footnote{
	This is a lower bound as we should assume that we did not find all the masquerading files.}.

If a ML or similarity-based solution is being used for security on files, then it is
important that the solution can effectively deal with masquerading files.
The use of digital signatures should be used as a part of ML solutions.
In addition, rulesets which test for stolen certificates, revoked certificates and certificates with a history
of signing malware should also be employed.
Digital signatures can be either applied as a post-processing step (as done in the various experiments in this paper)
or could be added as additional input features.

Special care should be used during the training and testing of ML systems.
ML systems should be tested with masquerading files to test whether they are capable of distinguishing between the legitimate versions and masquerading versions.
One approach to this may be to ensure that if legitimate software forms a part of the training set, then it is
important to include masquerading versions in the malicious part of the training set.

In conclusion,
\begin{itemize}
	\item Machine Learning and/or similarity should play an important role in security operations and security products
		for the identification of masquerading files.
	\item ML solutions in security need to consider how to deal with masquerading samples. Specifically,
		ML solutions for files need to be able to deal with masquerading files; one approach for this
		is to tightly couple the ML with a digital signature solution.
	\item Digital signatures are not a panacea. Many files are not signed, and further research needs to establish
		how ML solutions can effectively operate in the presence of masquerading samples for unsigned files.
		We encourage all software producers to adopt the use of digital signatures.
\end{itemize}

%
%
%
\bibliographystyle{plain}
\bibliography{references}

\end{document}